\documentclass[a4paper,11pt]{article}
\usepackage[in,plain]{fullpage}

\usepackage[colorlinks=true,allcolors=blue]{hyperref}

\usepackage{amsmath,amssymb}
\usepackage{mathptmx,cite,microtype}
\usepackage{graphicx}
\usepackage[small,bf]{caption}

\usepackage{tikz}
\usetikzlibrary{arrows.meta,decorations.pathmorphing}

\newcommand{\mku}{\ifmmode\mkern1mu\else0.167\thinspace\fi}
\renewcommand{\,}{\ifmmode\mkern2mu\else\thinspace\fi}

\newcommand{\PP}{\mathbb{P}}
\newcommand{\EE}{\mathbb{E}}

\renewcommand{\leq}{\ensuremath{\leqslant}}
\renewcommand{\geq}{\ensuremath{\geqslant}}

\DeclareMathOperator{\Real}{Re}
\DeclareMathOperator{\Imag}{Im}
\DeclareMathOperator*{\Res}{Res}

\makeatletter
\renewcommand\@makefntext[1]{\leftskip=0.0em\hskip-0.5em\@makefnmark#1}
\makeatother
\begin{document}

\renewcommand{\thefootnote}{\fnsymbol{footnote}}

\begin{titlepage}

\thispagestyle{empty}

\begin{center}

{\Large\textbf{Skewness tunes the small-drift record rate \\ of random walks and L\'{e}vy flights}}
\\[8ex]

{\large\textbf{Jos\'{e} Ricardo G. Mendon\c{c}a}\footnote{Email: {\tt\href{mailto:jricardo@usp.br}{\nolinkurl{jricardo@usp.br}}}.}}

\textit{\mbox{Escola de Artes, Ci\^{e}ncias e Humanidades, Universidade de S\~{a}o Paulo} \\ \mbox{Rua Arlindo Bettio 1000, Vila Guaraciaba, 03828-000 S\~{a}o Paulo, SP, Brazil}}
\\[8ex]

{\large\textbf{Abstract}}
\\[2ex]

\parbox{32em}
{A random walk with small positive drift $\mu$ sets new records at a rate $\lambda(\mu)$ that vanishes as $\mu \to 0$. For Gaussian and strictly stable centered steps whose stable law $Y$ has index $1 < \alpha \leq 2$ and positivity parameter $\rho = \PP(Y>0)$, we find $\lambda(\mu) \sim K\mu^{(1-\rho)/\nu}$ as $\mu \to 0$, where $\nu=1-1/\alpha$ and $K$ is explicit. Throughout their domains of attraction, the exponent persists, with a slowly varying factor replacing the constant $K$. The exponent is set by the asymmetry only through $\rho$, sweeping the interval $[1,\,1/(\alpha-1)]$ as the skewness varies. For centered strictly stable steps, $\rho$ also governs the driftless record growth, $\langle R_{N}\rangle \sim N^{\rho}/\Gamma(1+\rho)$, which the small-drift law meets at the crossover where the drift takes over. The formula recovers the Gaussian linear law $\lambda(\mu) \sim \sqrt{2}\mu/\sigma$ and, for symmetric heavy tails, the power $\mu^{\alpha/2(\alpha-1)}$. It follows from one Mellin transform of the harmonic sum in the Spitzer--Baxter identity, which factorizes into a kernel transform carrying the step distribution and a Riemann zeta function carrying the harmonic weights. Its poles deliver the leading law, its prefactor, and a correction ladder reproducing the known Gaussian and stable series, unifying diffusive, heavy-tailed, and skewed walks. The power law ends at the Cauchy point $\alpha=1$, where $\mu$ is a pure location shift: for the strictly Cauchy family the limiting rate vanishes and records accumulate sublinearly with a shift-dependent exponent.
\\

{\noindent}\textbf{Keywords}: \mbox{Record statistics} $\cdot$ \mbox{small-drift asymptotics} $\cdot$ \mbox{stable laws} $\cdot$ \mbox{Spitzer--Baxter identity} $\cdot$ \mbox{harmonic sums} $\cdot$ \mbox{Mellin transform}}

\end{center}

\end{titlepage}

%% %% %% %% %% %% %% %% %% %% %% %% %% %% %% %% %% %% %% %% %% %% %% %% %% %% %%

\section{Introduction}

Record statistics describe the advance of a running maximum and arise in hydrology, finance, evolutionary biology, and the physics of disordered and extreme systems \cite{WergenBogner2011,BrayMajumdar2013,Wergen2013,GodrecheMajumdar2017,SchehrMajumdar2014}. For symmetric random walks, the numbers and ages of records are {universal}, independent of the jump distribution, by the Sparre Andersen theorem \cite{SparreAndersen1954,MajumdarZiff2008}. The universality is robust: measurement error and noise rescale the pace of record setting but preserve the $\sqrt{n}$ growth \cite{Edery2013}. These laws, however, apply at zero drift. A drift breaks the universality: the record rate then depends on the step distribution, as for record temperatures under a warming trend \cite{Fischer2025}, and the question becomes how quickly that rate dies away as the drift is switched off.

\begin{figure}[!tb]
\centering
\includegraphics[width=0.63\columnwidth]{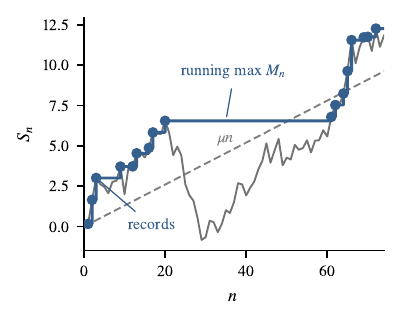} \\[-8pt]
\caption{\label{fig:setup}%
A biased random walk $S_{n}$ (grey) with small positive drift $\mu$, its running maximum $M_{n} = \max(S_{0},\dots,S_{n})$ (blue staircase), and its record epochs (dots). The mean grows as $\mu n$ (dashed). Records cluster near the running maximum and become sparse during long excursions below it. As $\mu \to 0$, the walk approaches its oscillating zero-drift regime and the record rate $\lambda(\mu)$ vanishes.}
\end{figure}

Let $\eta_{1},\eta_{2},\ldots$ be i.i.d.\ continuous random variables with finite mean zero. For a given $\mu>0$, set $\xi_{i}=\eta_{i}+\mu$. Then $\xi_{i}$ has mean $\mu$, and varying $\mu$ shifts the step distribution while preserving its shape. The random walk $S_{n}$, $n \geq 0$, is then
\begin{equation}
\label{eq:shiftfamily}
S_{0} = 0, \quad
S_{n} = S_{n-1} + \xi_{n} = \sum_{i=1}^{n}\xi_{i}.
\end{equation}
A {record} occurs at step $n$ when the walk exceeds all its earlier positions, and the successive times $0 = n_{0} < n_{1} < n_{2} < \cdots$ at which this happens are its record epochs, as illustrated in Figure~\ref{fig:setup}. The number of records through step $N$, including the initial record at $n = 0$, is
\begin{equation}
R_{N} = 1 + \#\{1 \leq n \leq N\colon S_{n} > \max(S_{0},\dots,S_{n-1})\}.
\end{equation}
The record rate is the long-run fraction of steps that set a record,
\begin{equation}
\label{eq:lambdadef}
\lambda(\mu) = \lim_{N \to \infty}\frac{R_{N}}{N}.
\end{equation}
At each record epoch the walk starts afresh from its current maximum, so the inter-record intervals are i.i.d.\ and the record epochs form a renewal process. The limit therefore exists and is nonrandom, equal to the reciprocal of the mean inter-record interval. The behavior of $\lambda(\mu)$ as $\mu \downarrow 0$ is governed by the fluctuations of the centered walk.

The small-drift behavior of the record rate reflects a competition of scales. Write $T_{n} = S_{n} - \mu n$ for the centered walk and suppose its steps are attracted to a stable law $Y$ of index $1 < \alpha \leq 2$, so that the fluctuations of $T_{n}$ grow like $n^{1/\alpha}$, up to a slowly varying correction. The drift grows linearly in $n$ while the fluctuations grow only as $n^{1/\alpha}$, so the two balance at a crossover $n^{*}$. When the attraction is {normal}, the fluctuation scale is $c\mku n^{1/\alpha}$ with a constant $c>0$, and
\begin{equation}
n^{*} \asymp (c/\mu)^{1/\nu}, \quad \nu = 1-1/\alpha,
\end{equation}
which recedes to infinity as $\mu \downarrow 0$. Below $n^{*}$ the walk is effectively driftless and persists above its starting point with probability of order $n^{-(1-\rho)}$, where $\rho = \PP(Y>0)$ is the positivity parameter of the limiting stable law: this is the Sparre Andersen--Doney persistence law \cite{SparreAndersen1954,Doney1995,BrayMajumdar2013}. Above $n^{*}$ the drift dominates and records accrue at a positive limiting rate \cite{WergenBogner2011,MajumdarSchehr2012}. Evaluating the driftless persistence probability at the crossover then suggests
\begin{equation}
\label{eq:heuristic}
\lambda(\mu) \asymp (n^{*})^{-(1-\rho)} \asymp \mu^{(1-\rho)/\nu}.
\end{equation}
The stable tail index thus fixes the crossover, whereas the positivity parameter carries the asymmetry.

Several strands of fluctuation theory and record statistics bear on the small-drift regime. For finite-variance walks, Spitzer's fluctuation identities and ladder-height theory give the linear small-drift law \cite{Spitzer1960,Feller1971}, and Chang and Peres obtained the complete Gaussian series \cite{ChangPeres1997}; Wergen, Bogner, and Krug later treated the same Gaussian regime in record-statistics terms, deriving a finite-time small-drift expansion and reporting the numerical limiting behavior $\lambda(\mu) \simeq 1.39\,\mu/\sigma$ \cite{WergenBogner2011}. For stable steps, Majumdar, Schehr, and Wergen mapped the phase diagram of biased records for symmetric jump densities, expressing the fixed-drift record rate as an exact Spitzer-type sum whose small-drift behavior was not extracted; in particular, the theory for asymmetric jump distributions is among the open problems they list \cite{MajumdarSchehr2012}. Skewness is covered, for strictly stable steps only, by the series of Hurvich and Reed for the probability that the all-time maximum of a negative-drift stable walk equals zero, an atom that reflection identifies with the record rate of the sign-reversed walk \cite{HurvichReed2016}. For general stable domains of attraction, Wachtel's transition theorem for ladder epochs gives the regular-variation form, but leaves the slowly varying multiplier implicit and does not provide a correction expansion \cite{Wachtel2009}.

In this paper we show how a common Mellin structure connects these results. The Spitzer--Baxter identity in Section~\ref{sec:harmonic} expresses the record rate of any continuous positive-drift walk as the exponential of a harmonic sum over the one-sided marginals $\PP(S_{n} \leq 0)$. For Gaussian and strictly stable increments these probabilities depend on $n$ and $\mu$ only through the combination $\mu n^{\nu}/c$. Section~\ref{sec:mellin} shows that the Mellin transform of the sum factorizes as a kernel transform carrying the step distribution times a Riemann zeta factor carrying the harmonic weights. Its poles organize the small-drift expansion. Related Mellin-pole methods have recently been applied to the first ascending ladder height of symmetric driftless walks \cite{GodrecheLuck2025}. In Section~\ref{sec:leading}, the double pole at the origin gives the exponent $(1-\rho)/\nu$ and the explicit prefactor $K$, with skewness entering the exponent through $\rho$. Beyond the exact kernels, reflection transfers Wachtel's theorem to the record rate, so the exponent persists throughout the stable domain of attraction as an index of regular variation. In the exact stable cases the leading law reads
\begin{equation}
\label{eq:main}
\lambda(\mu) \sim K\mu^{(1-\rho)/\nu}, \quad \mu \to 0,
\end{equation}
with the exponent swept across ${[1,\,1/(\alpha-1)]}$ as the skewness varies. Section~\ref{sec:cauchy} treats the singular boundary of this law at the strictly Cauchy point $\alpha=1$, where the limiting rate vanishes for every $\mu$ and the mean record number grows sublinearly with a shift-dependent exponent. The negative-integer poles generate the correction ladders of Chang and Peres and of Hurvich and Reed in Section~\ref{sec:ladder}, and Section~\ref{sec:numerics} tests the leading exponent across the stable family and the Gaussian correction ladder by deterministic evaluation of the Spitzer--Baxter sum. Section~\ref{sec:max} applies the factorization to the expected maximum. Shifting the Mellin variable by one unit places Kingman's heavy-traffic term \cite{Kingman1962} and Siegmund's corrected-diffusion constant \cite{Siegmund1979} at adjacent poles.

%% %% %% %% %% %% %% %% %% %% %% %% %% %% %% %% %% %% %% %% %% %% %% %% %% %% %%

\section{Record rate as a harmonic sum}
\label{sec:harmonic}

Because the increment distribution is continuous, ties occur with probability zero and the record epochs are the strict ascending ladder epochs. Let
\begin{equation}
\tau_{+} = \inf\{n \geq 1\colon S_{n}>0\}
\end{equation}
be the first strict ascending ladder epoch. Successive record intervals are independent copies of $\tau_{+}$, and the elementary renewal theorem gives
\begin{equation}
\label{eq:renewalrate}
\lambda(\mu) = \frac{1}{\EE(\tau_{+})}.
\end{equation}
Reversing the first $n$ increments identifies the probability that step $n$ is a record with the $n$-step survival probability. Taking $n \to \infty$ gives
\begin{equation}
\label{eq:survivalrate}
\lambda(\mu) = \PP(S_{n}>0 \text{ for every } n \geq 1).
\end{equation}
The Spitzer--Baxter identity \cite{Spitzer1956,Baxter1958,Feller1971} expresses this survival probability in terms of the one-sided marginals,
\begin{equation}
\label{eq:spitzerbaxter}
\lambda(\mu) = \exp\biggl[-\sum_{n=1}^{\infty}\frac{1}{n}\,\PP(S_{n} \leq 0)\biggr] = \exp[-g(\mu)],
\end{equation}
valid for any continuous i.i.d.\ step distribution, with $\lambda(\mu)=0$ when the sum diverges. A positive mean makes the sum converge, and then $\lambda(\mu)>0$.

The sum in \eqref{eq:spitzerbaxter} becomes a harmonic sum when the one-sided probabilities depend on $n$ and $\mu$ through a single scaling variable. Write $T_{n} = \sum_{i=1}^{n}\eta_{i}=S_{n}-\mu n$ and suppose first that $\eta_{i}$ is Gaussian or strictly stable. For $1 < \alpha < 2$, let $Y$ have the standard stable characteristic function \cite{Zolotarev1986}
\begin{equation}
\label{eq:cf}
\varphi_{\alpha,\beta}(t) = \exp\bigl[-|t|^{\alpha}\bigl(1-i\beta\tan{(\tfrac{1}{2}\pi\alpha)}\operatorname{sgn}t\bigr)\bigr],
\quad -1 \leq \beta \leq 1.
\end{equation}
In the Gaussian case, set $\alpha=2$ and take $Y$ to be standard normal. With an appropriate scale $c>0$, stability then gives the distributional identity
\begin{equation}
\label{eq:exactscaling}
T_{n} \,\stackrel{d}{ = }\, c\mku n^{1/\alpha}Y \quad (n \geq 1).
\end{equation}
Consequently, with $h(x) = \PP(Y \leq -x)$ and $\nu=1-1/\alpha$,
\begin{equation}
\label{eq:scaling}
\PP(S_{n} \leq 0) = \PP\Bigl(Y \leq -\frac{\mu n^{\nu}}{c}\Bigr) = h\Bigl(\frac{\mu n^{\nu}}{c}\Bigr).
\end{equation}
In these cases the Spitzer--Baxter exponent is exactly the harmonic sum
\begin{equation}
\label{eq:harmonic}
g(\mu) = \sum_{n=1}^{\infty}\frac{1}{n}\,h\Bigl(\frac{\mu n^{\nu}}{c}\Bigr).
\end{equation}
The drift enters only through the scaling variable $\mu n^{\nu}/c$, which is of order one at the crossover $n^{*}$.

For a general centered distribution in the domain of attraction of $Y$, stable convergence takes the form
\begin{equation}
\frac{T_{n}}{b_{n}} \Rightarrow Y, \quad b_{n} = n^{1/\alpha}\ell_{b}(n),
\end{equation}
and supplies a stable approximation to $\PP(S_{n} \leq 0)$ on the crossover scale. The small-$n$ part of \eqref{eq:spitzerbaxter}, the rate of convergence to the stable law, and the slowly varying function $\ell_{b}$ can all affect the prefactor and the lower-order terms. Accordingly, the Mellin calculation uses the exact Gaussian and strictly stable kernels, and the general regular-variation result comes from ladder-epoch theory.

%% %% %% %% %% %% %% %% %% %% %% %% %% %% %% %% %% %% %% %% %% %% %% %% %% %% %%

\section{Mellin factorization and pole structure}
\label{sec:mellin}

Harmonic sums like \eqref{eq:harmonic} factorize under the Mellin transform \cite{FlajoletGourdon1995}. Define
\begin{equation}
\label{eq:hstar}
h^{*}(s) = \int_{0}^{\infty}u^{s-1}h(u)\,du.
\end{equation}
The stable tail satisfies $h(u)=O(u^{-\alpha})$ as $u \to \infty$ when $1 < \alpha < 2$, while the Gaussian tail decreases faster than any power. At the other endpoint,
\begin{equation}
\label{eq:hsmall}
h(u) = 1-\rho + O(u), \quad u \downarrow 0.
\end{equation}
It follows that $h^{*}$ is analytic in $0<\Real{s}<\alpha$. For the Gaussian the right edge can be moved arbitrarily far to the right, as it can for maximally positively skewed steps, whose descending tail is exponentially thin rather than algebraic.

\begin{figure}[t]
\centering
\begin{tikzpicture}[
	x=1.40cm, y=1.40cm, >=Latex,
	every node/.style={font=\footnotesize},
	axis/.style={line width=0.9pt},
	contour/.style={line width=0.9pt},
	guide/.style={densely dashed, line width=0.6pt},
	pole/.style={line width=0.9pt},
	wave/.style={decoration={snake, amplitude=0.8mm, segment length=16mm}},
]
\def\cnot{1.35}
\def\alpharight{3.10}
%% Fundamental strip 0 < Re s < alpha
\fill[gray!16, wave]
   (0,1.95) -- (0,-1.20)
   decorate {-- (\alpharight,-1.20)}
   -- (\alpharight,1.95)
   decorate {-- (0,1.95)}
   -- cycle;
\draw[guide] (\alpharight,-1.20) -- (\alpharight,1.95);
\node[below] at (\alpharight,-1.20) {$\alpha$};
%% Axes
\draw[axis,->] (-4.40,0) -- (3.55,0) node[below=1pt] {$\Real{s}$};
\draw[axis,->] (0,-1.20) -- (0,2.30) node[right=1pt] {$\Imag{s}$};
%% Mellin inversion contour
\draw[contour] (\cnot,-1.16) -- (\cnot,1.94);
\node[below] at (\cnot,-1.20) {$a$};
%% Small-drift contour shift
\draw[guide,->] (\cnot-0.10,1.40) -- (-3.70,1.42);
\node[above] at (-1.65,1.50) {small-drift contour shift};
%% Correction poles at the negative integers
\foreach \x in {-1,-2,-3,-4} {
	\draw[pole] (\x-0.08,-0.08) -- (\x+0.08,0.08);
	\draw[pole] (\x-0.08,0.08) -- (\x+0.08,-0.08);
	\node[below=3pt] at (\x,0) {$\x$};
}
%% Double pole at the origin
\draw[pole] (-0.08,-0.08) -- (0.08,0.08);
\draw[pole] (-0.08,0.08) -- (0.08,-0.08);
\draw[pole] (0,0) circle (0.12);
\node[anchor=west] at (0.18,-0.24) {$s=0$};
%% Correction-ladder label
\node[above] at (-2.50,0.15) {$\zeta(1-\nu k)$ correction ladder};
%% Double-pole annotation
\node[align=center,anchor=east] (dp) at (-0.95,-1.05)
	{double pole \\ $\lambda(\mu) \sim K\mu^{(1-\rho)/\nu}$};
\draw[guide,->] (dp.east) .. controls (-0.55,-0.78) and (-0.22,-0.36) .. (-0.10,-0.13);
%% Fundamental-strip label
\node[fill=gray!16, inner sep=3pt] at (1.60,0.90) {fundamental strip};
\end{tikzpicture}
\caption{\label{fig:poles}Poles of
$G(s)=c^{s}h^{*}(s)\zeta(1+\nu s)$. A double pole at $s=0$ gives the leading
law, and simple poles at the negative integers give the correction ladder.
Moving the inversion contour to the left produces successive terms of the
small-drift expansion.}
\end{figure}
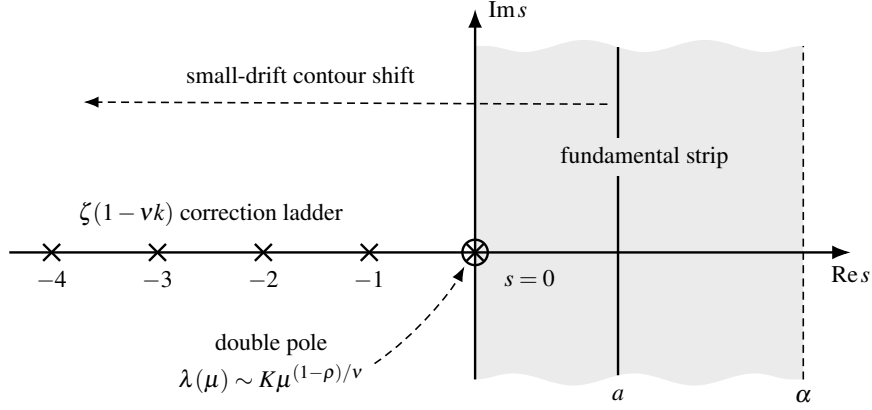

For $\Real{s}$ in this strip, positivity of $h$ and absolute convergence justify interchanging the sum and the integral. Substitution of $u = \mu n^{\nu}/c$ then gives
\begin{equation}
\label{eq:Gmaster}
%\begin{split}
G(s) = \int_{0}^{\infty} \mu^{s-1}g(\mu)\,d\mu % \\[1ex] 
= \sum_{n=1}^{\infty}\frac{1}{n} \int_{0}^{\infty}\mu^{s-1}h\Bigl(\frac{\mu n^{\nu}}{c}\Bigr)\,d\mu %\\[1ex] 
= c^{s}h^{*}(s)\sum_{n=1}^{\infty}n^{-1-\nu s} = c^{s}\,h^{*}(s)\zeta(1+\nu s).
%\end{split}
\end{equation}
Thus the lower-tail kernel contains the distributional information, while the Riemann $\zeta$ factor contains the harmonic weights. Mellin inversion recovers the Spitzer--Baxter exponent,
\begin{equation}
\label{eq:inversion}
g(\mu) = \frac{1}{2\pi i}\int_{a-i\infty}^{a+i\infty} G(s)\mu^{-s}\,ds, \quad 0 < a < \alpha.
\end{equation}
The small-$\mu$ expansion follows by continuing $G$ to the left and displacing the contour across its poles, as illustrated in Figure~\ref{fig:poles}. The behavior \eqref{eq:hsmall} continues $h^{*}$ across $s=0$ after subtracting $(1-\rho)/s$. Further Taylor coefficients of $h$ generate the poles at the negative integers.

For unit-variance Gaussian steps the kernel is $h(x)=\Phi(-x)=\tfrac{1}{2}\operatorname{erfc}(x/\sqrt{2})$ and the base transform is explicit. Integration by parts, with $h'(x)=-e^{-x^{2}/2}/\sqrt{2\pi}$ and boundary terms that vanish at both ends of the strip, followed by the substitution $t=x^{2}/2$, gives
\begin{equation}
\label{eq:hgaussbyparts}
h^{*}(s) = \frac{1}{s\sqrt{2\pi}}\int_{0}^{\infty}x^{s}e^{-x^{2}/2}\,dx = \frac{2^{(s-1)/2}}{s\sqrt{2\pi}}\int_{0}^{\infty}t^{(s-1)/2}e^{-t}\,dt.
\end{equation}
The last integral is $\Gamma(\frac{s+1}{2})$, so that
\begin{equation}
\label{eq:hgauss}
h^{*}(s) = \frac{2^{s/2-1}}{s\sqrt{\pi}}\,\Gamma\Bigl(\frac{s+1}{2}\Bigr).
\end{equation}
This expression is meromorphic in the whole plane, with simple poles at $s=0$ and at the negative odd integers only, where the Gamma factor is singular. Even poles are absent because the expansion of $\Phi(-x)$ about the origin contains, beyond the constant term, only odd powers of $x$. Expanding \eqref{eq:hgauss} about $s=0$ with $\psi(\tfrac{1}{2}) = -\gamma-2\log{2}$, where $\psi(z) = \Gamma'(z)/\Gamma(z)$ is the digamma function and $\gamma$ denotes the Euler-Mascheroni constant, gives the Laurent behavior
\begin{equation}
\label{eq:hgausslaurent}
h^{*}(s) = \frac{1}{2s}-\frac{\gamma+\log{2}}{4}+O(s),
\end{equation}
whose constant term reappears below as the finite part $h_{0}^{*}$ in the leading-law prefactor.

%% %% %% %% %% %% %% %% %% %% %% %% %% %% %% %% %% %% %% %% %% %% %% %% %% %% %%

\section{Leading law and skew-tuned exponent}
\label{sec:leading}

\subsection{The double pole}

Let the centered increments be Gaussian or strictly stable with index
$1<\alpha \leq 2$, scale $c>0$, positivity parameter $\rho=\PP(Y>0)$, and
$\nu=1-1/\alpha$. The finite part of the base transform at the origin is
\begin{equation}
\label{eq:hstarfinite}
h_{0}^{*} = \int_{0}^{1}\frac{h(u)-(1-\rho)}{u}\,du + \int_{1}^{\infty}\frac{h(u)}{u}\,du.
\end{equation}
The local behavior \eqref{eq:hsmall} gives the meromorphic continuation
\begin{equation}
\label{eq:laurent}
h^{*}(s) = \frac{1-\rho}{s} + h_{0}^{*} + O(s).
\end{equation}
At the same point,
\begin{equation}
\zeta(1+\nu s) = \frac{1}{\nu s} + \gamma + O(s),
\quad
c^{s} = 1 + s\log{c} + O(s^{2}).
\end{equation}
Multiplying the three expansions above yields
\begin{equation}
\label{eq:Glaurent}
G(s) = \frac{1-\rho}{\nu s^{2}} + \frac{P}{s} + O(1),
\end{equation}
where
\begin{equation}
\label{eq:Pgeneral}
P = (1-\rho)\gamma + \frac{h_{0}^{*}}{\nu} + \frac{1-\rho}{\nu}\log{c}.
\end{equation}
The stable density is $C^{\infty}$ and, for $1 < \alpha < 2$, its $j$-th derivative satisfies $f_{Y}^{(j)}(-u) = O(u^{-\alpha-1-j})$ as $u \to \infty$. In the Gaussian case all its derivatives decrease faster than any power \cite{Zolotarev1986}. Integration by parts gives
\begin{equation}
\label{eq:hbyparts}
h^{*}(s) = \frac{1}{s}\int_{0}^{\infty} u^{s} f_{Y}(-u)\,du.
\end{equation}
After the substitution $u=e^{x}$, the integral is the Fourier transform of a smooth function whose derivatives are integrable uniformly on compact substrips of $-1 < \Real{s} < \alpha$. Thus, for every $M$, $h^{*}(s) = {O_{M}((1+|\Imag{s}|)^{-M})}$ on such substrips. Since $\zeta(1+\nu s)$ grows at most polynomially on vertical lines \cite{FlajoletGourdon1995}, the inversion integral is absolutely convergent on each shifted line and the horizontal sides of the contour vanish. We may therefore, for any fixed $0 < \delta < 1$, move the inversion line in \eqref{eq:inversion} from $\Real{s}=a$ to $\Real{s} = -\delta$. The only crossed singularity is the double pole at $s=0$, and
\begin{equation}
\Res_{s=0}\bigl\{G(s)\mu^{-s}\bigr\} = \frac{1-\rho}{\nu}\log{\frac{1}{\mu}} + P.
\end{equation}
The integral on the shifted line is $O(\mu^{\delta})$, because $|\mu^{-s}|= \mu^{\delta}$ there. Hence
\begin{equation}
g(\mu) = \frac{1-\rho}{\nu}\log{\frac{1}{\mu}} + P + O(\mu^{\delta}),
\end{equation}
so that
\begin{equation}
\label{eq:leadinglaw}
\lambda(\mu) = K\mu^{(1-\rho)/\nu}\bigl[1+O(\mu^{\delta})\bigr],
\quad \mu \downarrow 0,
\end{equation}
with $K=e^{-P}$.

\subsection{General domains of attraction}

Suppose the continuous centered increment distribution belongs to the domain of attraction of a stable law $Y$ with index $1<\alpha \leq 2$ and positivity parameter $\rho = \PP(Y>0)$. The reflected centered increment $-\eta$ has a stable limit with positivity parameter $1-\rho$, and $\tau_{+}$ is the first descending ladder epoch of the reflected walk with mean increment $-\mu$. Theorem~3 of Wachtel \cite{Wachtel2009}, whose stable convention coincides with \eqref{eq:cf}, applied with moment order $r=1$, gives, for some slowly varying function $\ell_{\lambda}$,
\begin{equation}
\EE(\tau_{+}) = \frac{\mu^{-\alpha(1-\rho)/(\alpha-1)}}{\ell_{\lambda}(1/\mu)}.
\end{equation}
The admissibility condition of that theorem reads $\rho<1<\alpha$ here, and holds throughout the range considered. Taking the reciprocal in \eqref{eq:renewalrate} and using $\nu=(\alpha-1)/\alpha$ gives, as $\mu \downarrow 0$,
\begin{equation}
\label{eq:regularvariation}
\lambda(\mu) = \mu^{(1-\rho)/\nu}\,\ell_{\lambda}(1/\mu).
\end{equation}
Thus Wachtel's result identifies $(1-\rho)/\nu$ as the index of regular variation throughout the stable domain of attraction, while $\ell_{\lambda}$ carries the remaining distributional information. At this level of generality $\ell_{\lambda}$ need not tend to a constant. For the Gaussian and strictly stable kernels treated above, the Mellin calculation gives $\ell_{\lambda}(1/\mu) \to K$.

\subsection{Dependence on skewness}

\begin{figure}[!t]
\centering
\includegraphics[width=0.62\columnwidth]{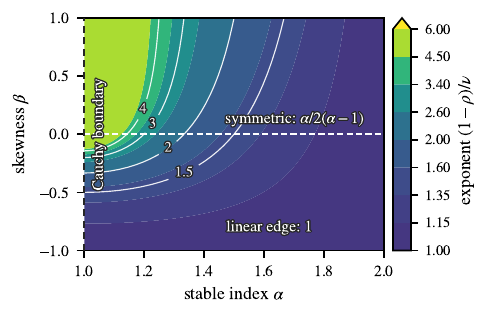} \\[-8pt]
\caption{\label{fig:phase}%
Phase portrait of the small-drift record rate. The leading exponent $(1-\rho)/\nu$ of $\lambda(\mu) \sim K\mu^{(1-\rho)/\nu}$ over the stable index ${1 < \alpha \leq 2}$ and skewness $-1 \leq \beta \leq 1$, from Eqs.~\eqref{eq:zolotarev}--\eqref{eq:exprange}. It equals $1$ along the entire diffusive edge $\alpha = 2$ and the descending-skew edge $\beta = -1$, and rises to $1/(\alpha-1)$ on the ascending-skew edge $\beta = +1$. The surface becomes singular as $\alpha \to 1^{+}$. The strictly Cauchy boundary, where the small-drift scaling changes character, is treated in Section~\ref{sec:cauchy}.}
\end{figure}

For $1 < \alpha < 2$, in the convention of \eqref{eq:cf}, the positivity parameter is the Zolotarev value \cite{Doney1995,Zolotarev1986}
\begin{equation}
\label{eq:zolotarev}
\rho = \frac{1}{2} + \frac{1}{\pi\alpha}
\arctan{(\beta\tan{(\tfrac{1}{2}\pi\alpha)})}
\in \Bigl[1-\frac{1}{\alpha},\,\frac{1}{\alpha}\Bigr],
\end{equation}
where the arctangent is taken in its principal branch. Therefore,
\begin{equation}
\label{eq:exprange}
\frac{1-\rho}{\nu}
\in \Bigl[1,\,\frac{1}{\alpha-1}\Bigr]
\end{equation}
as $\beta$ runs from $-1$ to $+1$. The ordering is counterintuitive, and
centering explains it: a heavier ascending tail might seem
to favor records, but at zero mean it is offset by a left-shifted bulk, so
the median lies below the mean, $\rho=\PP(Y>0)$ falls below $1/2$, and the
persistence that carries record times decays faster. Accordingly, at
$\beta=-1$ the heavier tail is on the descending side, $\rho=1/\alpha$ is
largest, and the exponent takes its smallest value $1$; the light ascending
tail then controls the ladder structure that carries records. At $\beta=+1$
the heavier tail is on the ascending side, $\rho=1-1/\alpha$ is smallest,
and the exponent reaches its maximum $1/(\alpha-1)$, so records are
suppressed most strongly.
For symmetric steps the dependence is absent: $\rho=1/2$ at every $\alpha$,
and the exponent collapses to the single value $\alpha/2(\alpha-1)$.
Figure~\ref{fig:phase} displays the full dependence.

For Gaussian increments with variance $\sigma^{2}$, one has $\rho=\nu=1/2$,
$c=\sigma$, and, from \eqref{eq:hgausslaurent},
$h_{0}^{*}=-(\gamma+\log{2})/4$. Equation \eqref{eq:Pgeneral} then gives
$P=\log{\sigma}-\tfrac{1}{2}\log{2}$, and \eqref{eq:leadinglaw} yields
\begin{equation}
\label{eq:sqrttwo}
\lambda(\mu) \sim \sqrt{2}\,\frac{\mu}{\sigma}.
\end{equation}
By \eqref{eq:renewalrate} and Wald's identity applied to the ascending ladder height $H_{+}=S_{\tau_{+}}$, one has $\EE_{\mu}(H_{+}) = \mu\,\EE(\tau_{+})$ and hence $\lambda(\mu) = \mu/\EE_{\mu}(H_{+})$. The coefficient in \eqref{eq:sqrttwo} is therefore the reciprocal of the mean driftless ascending ladder height, $\EE_{0}(H_{+}) = \sigma/\sqrt{2}$ \cite{ChangPeres1997,Spitzer1960}. It is the exact value behind the numerical estimate $\lambda(\mu) \simeq 1.39\,\mu/\sigma$ of \cite{WergenBogner2011}: at finite drift the ratio $\lambda(\mu)/\mu$ approaches $\sqrt{2}/\sigma$ from below through the correction ladder of Section~\ref{sec:ladder}, as Table~\ref{tab:gausscheck} shows. More generally, every continuous finite-variance centered distribution has the linear exponent. Its coefficient is $\sqrt{2}/(\kappa\sigma)$ and depends on the increment distribution. Writing $S_{k}^{(0)} = \eta_{1} + \cdots + \eta_{k}$,
\begin{equation}
\label{eq:spitzerconstant}
\lambda(\mu) \sim \frac{\sqrt{2}}{\kappa\sigma}\mku\mu,
\quad
\kappa = \exp\biggl\{\sum_{k=1}^{\infty}\frac{1}{k} \Bigl[\frac{1}{2}-\PP(S_{k}^{(0)}>0)\Bigr]\biggr\}.
\end{equation}
The series converges for every centered finite-variance distribution, by the Spitzer--Ros\'{e}n theorem \cite{Spitzer1960,Rosen1962}. For symmetric increments every probability in brackets equals $1/2$, so $\kappa=1$.

For symmetric strictly stable increments, $\rho=1/2$ and the natural scale $c=1$ gives
\begin{equation}
\label{eq:symheavy}
\lambda(\mu) \sim \mu^{\alpha/2(\alpha-1)}, \quad 1 < \alpha < 2.
\end{equation}
The exponent exceeds $1$ and decreases to the diffusive value as $\alpha \uparrow 2$.

\subsection{Explicit strictly stable prefactor}

For $1 < \alpha < 2$, let $f_{Y}$ be the density of the standard stable law in \eqref{eq:cf}. Integration by parts in Eqs.~\eqref{eq:hstar} and \eqref{eq:hbyparts} gives
\begin{equation}
\label{eq:hstarlog}
h_{0}^{*} = \int_{0}^{\infty}(\log{u})f_{Y}(-u)\,du = \EE[\log{|Y|};Y{<0}].
\end{equation}
Thus the constant term is a one-sided logarithmic moment. Set
\begin{equation}
R_{\alpha,\beta} = \bigl|1-i\beta\tan{(\tfrac{1}{2}\pi\alpha)}\bigr|.
\end{equation}
For the convention \eqref{eq:cf}, the one-sided fractional-moment identity is given in \cite{Zolotarev1986} as
\begin{equation}
\EE[|Y|^{s};Y{<0}] = R_{\alpha,\beta}^{s/\alpha}\, 
\frac{\Gamma(s)\Gamma(1-s/\alpha)}{\pi}\sin{(\pi(1-\rho)s)}, 
\quad -1 < \Real{s} < \alpha,
\end{equation}
with the value at $s = 0$ understood by continuity. The apparent singularity there is removable. Differentiating at $s = 0$ and using \eqref{eq:hstarlog} gives
\begin{equation}
\label{eq:hstar0skew}
h^{*}_{0} = (1-\rho)\Bigl[\gamma\Bigl(\frac{1}{\alpha}-1\Bigr) + \frac{1}{\alpha}\log{R_{\alpha,\beta}}\Bigr].
\end{equation}
Substitution in \eqref{eq:Pgeneral} yields
\begin{equation}
\label{eq:Kskew}
P = \frac{1-\rho}{\nu}\log{\Bigl(c\mku R_{\alpha,\beta}^{1/\alpha}\Bigr)},
\quad 
K = \Bigl(c\mku R_{\alpha,\beta}^{1/\alpha}\Bigr)^{-(1-\rho)/\nu}.
\end{equation}
For $\beta=0$ this reduces to $K = c^{-(1-\rho)/\nu}$ and hence to $K=1$ in the natural scale $c=1$. At $\alpha=3/2$ the spectrally one-sided endpoints are $K = 2^{-1/3}$ for $\beta=-1$ and $K = 2^{-2/3}$ for $\beta=+1$. The skewness thus acts on both parts of the leading law, but unequally: in the prefactor it enters only through $R_{\alpha,\beta}^{1/\alpha}$, which at $\alpha=3/2$ ranges over $[1,\,2^{1/3}]$, whereas in the exponent it sweeps the whole of \eqref{eq:exprange}, from $1$ to $1/(\alpha-1)=2$.

After converting Hurvich and Reed's stable parameterization to \eqref{eq:cf}, these strictly stable constants and exponents coincide with those of their all-time-maximum expansion \cite{HurvichReed2016}. The connection with the record rate follows by reflection,
\begin{equation}
\label{eq:reflection}
\lambda(\mu) = \PP(S_{n}>0 \text{ for every } n \geq 1) = \PP\Bigl(\sup_{n \geq 1}(-S_{n}) \leq 0\Bigr).
\end{equation}
The $s=0$ pole therefore recovers their result and displays its dependence on the positivity parameter. For a general domain of attraction, the slowly varying factor in \eqref{eq:regularvariation} carries the distributional dependence.

%% %% %% %% %% %% %% %% %% %% %% %% %% %% %% %% %% %% %% %% %% %% %% %% %% %% %%

\section{The Cauchy point and the driftless law}
\label{sec:cauchy}

The phase portrait has a singular boundary at $\alpha=1$. At this boundary the increment mean does not exist; $\mu$ is the location parameter of the shifted Cauchy law. The scale separation behind \eqref{eq:scaling} requires $\nu = 1-1/\alpha>0$. As $\alpha \downarrow 1$, the crossover $n^{*} \asymp (c/\mu)^{1/\nu}$ recedes faster than any fixed power of $1/\mu$. Exactly at $\alpha=1$ the accumulated shift $\mu n$ and the fluctuations both grow linearly with $n$, so there is no analogous crossover.

Consider symmetric Cauchy increments with location zero and scale $c$, shifted by $\mu$. Then $S_{n}/n$ is Cauchy with location $\mu$ and scale $c$ for every $n$, and
\begin{equation}
\label{eq:cauchypos}
\PP(S_{n} \leq 0) = \frac{1}{2}-\frac{1}{\pi}\arctan{\Bigl(\frac{\mu}{c}\Bigr)}
\end{equation}
is independent of $n$. The harmonic sum in \eqref{eq:spitzerbaxter} therefore diverges logarithmically and
\begin{equation}
\lambda(\mu) = 0
\end{equation}
for every finite $\mu$. The number of records grows sublinearly.

The finite-time growth follows from the corresponding finite Spitzer generating function. Let $q_{n}$ be the probability that step $n$ is a record, with $q_{0}=1$, and let
\begin{equation}
\theta(\mu) = \PP(S_{n}>0) = \frac{1}{2} + \frac{1}{\pi}\arctan{\Bigl(\frac{\mu}{c}\Bigr)}
\end{equation}
be the positivity probability of the drifted walk. Because it is independent of $n$,
\begin{equation}
\label{eq:cauchygen}
%\begin{split}
\sum_{n=0}^{\infty}q_{n}\mku z^{n} 
= \exp\biggl[\sum_{n=1}^{\infty}\frac{z^{n}}{n}\PP(S_{n}>0)\biggr] %\\[1ex] 
= \exp\biggl[\theta(\mu)\sum_{n=1}^{\infty}\frac{z^{n}}{n}\biggr] 
= (1-z)^{-\theta(\mu)}.
%\end{split}
\end{equation}
Taking coefficients gives the exact finite-$n$ probability
\begin{equation}
q_{n} = \frac{\Gamma(n+\theta(\mu))}{\Gamma(\theta(\mu))\Gamma(n+1)} \sim \frac{n^{\theta(\mu)-1}}{\Gamma(\theta(\mu))}.
\end{equation}
Summing $q_{n}$ up to $N$ yields
\begin{equation}
\label{eq:cauchyrec}
\langle R_{N}\rangle \sim \frac{N^{\theta(\mu)}}{\Gamma(1+\theta(\mu))},
\end{equation}
the strictly Cauchy result first obtained by Le Doussal and Wiese \cite{LeDoussalWiese2009} and extended across the $\alpha=1$ class by Majumdar, Schehr, and Wergen \cite{MajumdarSchehr2012}. See also \cite{RadiceCristadoro2025}. As $\nu \to 0$ in \eqref{eq:Gmaster}, the scaling variable $\mu n^{\nu}/c$ loses its $n$-dependence, tending to $\mu/c$ for fixed $n$. The kernel freezes, and the zeta factor remains at its pole. The power law in $\mu$ is replaced by a family of finite-time exponents $\theta(\mu)$. At zero shift the exponent is $\theta(0)=1/2$, the universal Sparre Andersen value; at positive shift the ratio $\mu/c$ enters the exponent itself. Equations \eqref{eq:cauchypos}--\eqref{eq:cauchyrec} are exact for the strictly Cauchy family. More general Cauchy domains require their own centering and slowly varying analysis \cite{Berger2019}.

The step that produced \eqref{eq:cauchygen} used only that $\PP(S_{n}>0)$ does not depend on $n$. That holds for the shifted Cauchy walk, and it holds just as well for a centered strictly stable walk of any index: for $1<\alpha \leq 2$, strict stability gives $S_{n} \stackrel{d}{=} c\mku n^{1/\alpha}Y$ at $\mu=0$, so $\PP(S_{n}>0) = \rho$ for every $n$. Equation \eqref{eq:cauchygen} applies with $\theta(\mu)$ replaced by $\rho$, and the driftless record number grows as
\begin{equation}
\label{eq:driftlessrho}
\langle R_{N}\rangle \sim \frac{N^{\rho}}{\Gamma(1+\rho)},
\quad \mu = 0, \quad 1 < \alpha \leq 2.
\end{equation}
For symmetric steps $\rho=1/2$ and \eqref{eq:driftlessrho} is the universal law $\langle R_{N}\rangle \sim 2\sqrt{N/\pi}$ \cite{SparreAndersen1954,MajumdarZiff2008}, which \eqref{eq:cauchyrec} also returns at $\mu=0$. For skewed steps it is not universal, and the skewness is already visible without any drift, in the growth exponent itself. For a general centered distribution in the domain of attraction of $Y$, Spitzer's condition $\PP(S_{n}>0) \to \rho$ replaces the exact identity and leaves the exponent unchanged up to a slowly varying factor.

For the strictly stable walks considered above, the driftless and small-drift laws agree at the crossover. For small $\mu$ and $N$ below $n^{*}$, the walk is effectively driftless and \eqref{eq:driftlessrho} applies. Above $n^{*}$, records accumulate linearly, $\langle R_{N}\rangle \simeq \lambda(\mu)N$. At $N \asymp n^{*} \asymp (c/\mu)^{1/\nu}$ the two carry the same power of the shift,
\begin{equation}
\label{eq:crossovermatch}
(n^{*})^{\rho} \asymp \lambda(\mu)\mku n^{*} \asymp \mu^{-\rho/\nu},
\end{equation}
so $(1-\rho)/\nu$ is the exponent that joins the linear law to the driftless one. In general domains of attraction, slowly varying factors modify the crossover scale and the leading multipliers but not this balance of powers.

%% %% %% %% %% %% %% %% %% %% %% %% %% %% %% %% %% %% %% %% %% %% %% %% %% %% %%

\section{Correction ladder}
\label{sec:ladder}

For exact Gaussian and strictly stable kernels, $1 < \alpha \leq 2$, the correction ladder follows from the local Taylor coefficients of $h$. General domains of attraction can have additional corrections governed by convergence to the stable law.

For any fixed $M$, smoothness of the stable density gives the local expansion
\begin{equation}
\label{eq:htaylor}
h(u) = \sum_{k=0}^{M}a_{k} u^{k}+O(u^{M+1}),
\quad
a_{0} = 1-\rho, \quad a_{1} = -f_{Y}(0).
\end{equation}
Subtract the Taylor polynomial after multiplying it by a smooth cutoff that equals one near the origin and vanishes for large $u$. The remainder is $O(u^{M+1})$ at the origin and retains the original tail at infinity. Its Mellin transform is therefore analytic in $-M-1<\Real{s}<\alpha$. Standard Mellin subtraction then gives
\begin{equation}
\label{eq:hstarpoles}
h^{*}(s) = \sum_{k=0}^{M}\frac{a_{k}}{s+k} + H_{M}(s),
\end{equation}
where $H_{M}$ is analytic throughout that strip. The derivative bounds used above, now applied to the cutoff-subtracted kernel, also give rapid decay on closed vertical substrips away from the poles. Thus $h^{*}$ has a possible simple pole at $s=-k$ with residue $a_{k}$. Since the zeta factor is regular there, the corresponding residue of the Mellin integrand is
\begin{equation}
\label{eq:correction}
\Res_{s=-k}\bigl\{G(s)\mu^{-s}\bigr\} = \frac{a_{k}}{c^{k}}\mku\zeta(1-\nu k)\mku\mu^{k}.
\end{equation}
For any fixed $0<\delta<1$, move the contour to $\Real{s} = -M-\delta$. The only possible crossed singularities are $s=0,-1,\ldots,-M$. The horizontal integrals vanish by the uniform vertical decay, and the integral on the new line is $O(\mu^{M+\delta})$. Hence
\begin{equation}
\label{eq:gseriesfixed}
g(\mu) = \frac{1-\rho}{\nu}\log{\frac{1}{\mu}} + P + \sum_{k=1}^{M}\frac{a_{k}}{c^{k}}\zeta(1-\nu k)\mu^{k} + O(\mu^{M+\delta}).
\end{equation}
Zeros of $\zeta(1-\nu k)$ and vanishing Taylor coefficients may remove individual terms. In particular,
\begin{equation}
\label{eq:lambdacorrection}
\lambda(\mu) = K\mu^{(1-\rho)/\nu} \Bigl[1-\frac{a_{1}}{c}\zeta(1-\nu)\mu+O(\mu^{2})\Bigr].
\end{equation}
Since $a_{1} = -f_{Y}(0)<0$ and $\zeta(1-\nu) < 0$ for $0 < \nu \leq 1/2$, that is, for $1 < \alpha \leq 2$, the bracket in \eqref{eq:lambdacorrection} approaches one from below.

For the symmetric stable kernel the even coefficients vanish and the odd ones are
\begin{equation}
\label{eq:sym-ak}
a_{2m+1} = -(-1)^{m}\,\frac{\Gamma((2m+1)/\alpha)}{\pi\alpha(2m+1)!},
\end{equation}
in the natural convention \eqref{eq:cf}, which at $\alpha=2$ normalizes the Gaussian to $\mathcal{N}(0,2)$. For a unit-variance Gaussian kernel, the coefficients are
\begin{equation}
\label{eq:gauss-ak}
a_{2m+1}^{\rm G} = -\frac{(-1)^{m}}{\sqrt{2\pi}\,2^{m}\,m!(2m+1)}.
\end{equation}
The two families differ at $\alpha=2$ by the factors $2^{k/2}$ between the normalizations. With variance $\sigma^{2}$, the analytic continuation of Chang and Peres \cite{ChangPeres1997} shows that the resulting series 
\begin{equation}
\label{eq:gaussseries}
g(\mu) = \log{\Bigl(\frac{\sigma}{\mu}\Bigr)} - \frac{1}{2}\log{2} + \sum_{m \geq 0}a_{2m+1}^{\rm G}\zeta(\tfrac{1}{2}-m)\Bigl(\frac{\mu}{\sigma}\Bigr)^{2m+1}
\end{equation}
converges for $0 < \mu/\sigma < 2\sqrt{\pi}$. In particular,
\begin{equation}
\label{eq:gausscorr}
\lambda(\mu) = \sqrt{2}\,\frac{\mu}{\sigma}\exp\biggl[\frac{\zeta(\frac{1}{2})}{\sqrt{2\pi}}\mku\frac{\mu}{\sigma} + O((\mu/\sigma)^{3})\biggr].
\end{equation}
For a non-Gaussian strictly stable kernel, \eqref{eq:gseriesfixed} is the Hurvich--Reed ladder \cite{HurvichReed2016}, since their zeta argument $k/\alpha-k+1$ equals $1-\nu k$. Godr\`{e}che and Luck derive related Mellin-pole expansions for the first ascending ladder height of symmetric Gaussian walks and L\'{e}vy flights, as part of a broader study of its distribution, moments, and asymptotic tail behavior \cite{GodrecheLuck2025}.

\section{Numerical verification of the pole expansion}
\label{sec:numerics}

For strictly stable increments the one-sided probability in the Spitzer--Baxter sum is known at every $n$, so the pole expansion can be checked by direct evaluation, free of sampling error. At scale $c=1$, writing $F_{Y}$ for the stable distribution function, we truncate the sum at $N$ and treat the remainder as an integral. Since $\mu n^{\nu}$ is the scaling variable, $dn/n = du/(\nu u)$ with $u=\mu n^{\nu}$, and
\begin{equation}
\label{eq:gnum}
g_{N}(\mu) = \sum_{n=1}^{N}\frac{1}{n}\mku F_{Y}(-\mu n^{\nu}) + \frac{1}{\nu}\int_{\mu N^{\nu}}^{\infty}\frac{F_{Y}(-u)}{u}\,du - \frac{F_{Y}(-\mu N^{\nu})}{2N},
\end{equation}
the last term being the Euler--Maclaurin endpoint correction. The CDF is tabulated on $3000$ uniformly spaced points of $[-60,0]$ by the Gil--Pelaez inversion formula applied to \eqref{eq:cf}, using $8192$-point Gauss--Legendre quadrature over $0 \leq t \leq 25$ and monotone cubic interpolation between grid points. The integral is evaluated from this tabulation up to the baseline cutoff $y=40$ and from its leading tail form beyond that point. Since $F_{Y}(-u) \sim C_{-}u^{-\alpha}$, the latter contributes $C_{-}y^{-\alpha}/(\alpha\nu)$ to $g_{N}(\mu)$; here $C_{-} = \lim_{x \to \infty}x^{\alpha}F_{Y}(-x)$. At the light-descending-tail endpoint $\beta=+1$, one has $C_{-}=0$ and the residual tail is exponentially small.

Table~\ref{tab:gausscheck} tests the unit-variance expansion through the cubic term. The discrepancy between the harmonic sum and \eqref{eq:gaussseries} scales as $\mu^{5}$, the first omitted odd power, while $\lambda(\mu)/\mu$ approaches $\sqrt{2}$ from below as predicted by
\eqref{eq:gausscorr}, consistent with the estimate $1.39$ of \cite{WergenBogner2011}.

\begin{table}[!t]
\centering
\renewcommand{\arraystretch}{1.10}
\begin{tabular}{cccc}
\hline
$\mu$ & $g(\mu)$ summed & series through $\mu^{3}$ & $\lambda(\mu)/\mu$\\
\hline
$0.10$ & $2.0142574$ & $2.0142574$ & $1.33419$\\
$0.05$ & $2.6782868$ & $2.6782868$ & $1.37361$\\
$0.02$ & $3.5771012$ & $3.5771012$ & $1.39783$\\
$0.01$ & $4.2644226$ & $4.2644226$ & $1.40600$\\
\hline
\end{tabular}
\caption{\label{tab:gausscheck}Gaussian correction ladder. Before rounding, the differences between the summed and truncated-series values are $3\times10^{-9}$, $8\times10^{-11}$, $8\times10^{-13}$, and $3\times10^{-14}$, respectively.}
\end{table}

For the numerically most demanding case, $\alpha=1.3$ and $\beta=0$, the fitted exponent changes by less than $10^{-7}$ between $N=10^{5}$ and $N=10^{7}$, and by less than $10^{-8}$ when the cutoff $y$ is moved from $20$ to $60$. The tabulation of $F_{Y}$ is then the leading error: halving the CDF grid spacing shifts the exponent by $6 \times 10^{-6}$. Using only the leading-power tail is inaccurate at small $\alpha$: at $N=10^{6}$ it gives $p=3.041$ for $\alpha=1.3$ and $p=1.418$ for $\alpha=3/2$, in place of $13/6$ and $3/2$.

\begin{figure}[!t]
\centering
\includegraphics[width=0.52\columnwidth]{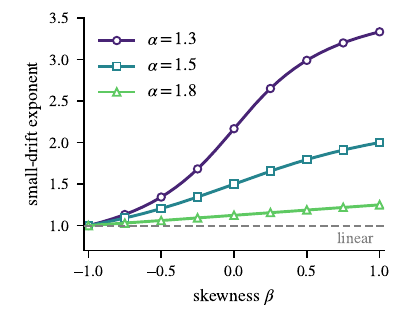} \\[-8pt]
\caption{\label{fig:exponent}%
Skew-tuned exponent. The leading record-rate exponent ${(1-\rho)/\nu}$ as a function of skewness, sweeping $[1,\,1/(\alpha-1)]$ as $\beta$ runs from $-1$ to $+1$ for fixed $\alpha$. Curves are the prediction, Eqs.~\eqref{eq:zolotarev}--\eqref{eq:exprange}. Open symbols are the exponents fitted to deterministic evaluations of the Spitzer--Baxter sum~\eqref{eq:spitzerbaxter}, as in Section~\ref{sec:numerics}; they follow the curves to better than $10^{-4}$ at every $(\alpha,\beta)$ shown. The dashed line marks the linear small-drift exponent, $(1-\rho)/\nu=1$. Exponents above it suppress records more strongly for small $\mu$.}
\end{figure}

To estimate the exponent independently of the correction coefficients, we fit
\begin{equation}
\log{\lambda(\mu)} = A+p\log{\mu}+B\mu+C\mu^{2}
\end{equation}
to the sums \eqref{eq:gnum} at $N=10^{6}$ on a logarithmic grid of twelve points spanning $0.02 \leq \mu \leq 0.2$. Repeating the fit for $\alpha=1.3$, $3/2$, and $1.8$ and for the nine skewnesses $\beta = -1, -3/4, \ldots, +1$ gives a largest discrepancy of $7 \times 10^{-5}$ from \eqref{eq:exprange} over the twenty-seven cases. Figure~\ref{fig:exponent} displays the fitted values against the predicted curves. For $\beta=0$, the fitted values are $p=2.1667$, $1.5000$, and $1.1250$, compared with $13/6$, $3/2$, and $9/8$ from \eqref{eq:symheavy}. At the spectrally one-sided endpoints, the fitted exponent ranges from $p=1.0000$ at $\beta=-1$ to $p=3.3333$ at $\alpha=1.3$, $\beta=+1$, compared with $10/3$. No extrapolation in $N$ is used.

%% %% %% %% %% %% %% %% %% %% %% %% %% %% %% %% %% %% %% %% %% %% %% %% %% %% %%

\section{Expected maximum}
\label{sec:max}

The expected maximum, a classical functional of drifted walks and L\'{e}vy flights \cite{ComtetMajumdar2005,MajumdarComtet2006,MounaixMajumdar2018}, can also be expressed through the base transform $h^{*}$. Let $M_{n} = \max(S_{0},\ldots,S_{n})$, and write $x^{+}=\max(x,0)$ and $x^{-}=\max(-x,0)$ for the positive and negative parts, so that $x=x^{+}-x^{-}$ and both are nonnegative. Spitzer's identity gives \cite{Spitzer1956,Feller1971}
\begin{equation}
\label{eq:spitzermax}
\EE(M_{N}) = \sum_{n=1}^{N}\frac{1}{n}\EE(S_{n}^{+}).
\end{equation}
Since $\EE(S_{n})=n\mu$ and $S_{n}^{+}=S_{n}+S_{n}^{-}$,
\begin{equation}
\EE(M_{N})-N\mu = \sum_{n=1}^{N}\frac{1}{n}\EE(S_{n}^{-}).
\end{equation}
When the limit is finite, we write
\begin{equation}
\label{eq:Ddef}
D(\mu) = \lim_{N \to \infty}\bigl[\EE(M_{N})-N\mu\bigr] = \sum_{n=1}^{\infty}\frac{1}{n}\EE(S_{n}^{-}).
\end{equation}
We call $D(\mu)$ the limiting excess, since the expected maximum runs ahead of the mean displacement $N\mu$ by this amount. Applying \eqref{eq:spitzermax} to the reflected walk $-S$, whose drift is $-\mu<0$, turns the sum into $\EE[\max_{k \leq N}(-S_{k})]$, so that
\begin{equation}
\label{eq:Dsup}
D(\mu) = \EE\Bigl[\sup_{n \geq 0}(-S_{n})\Bigr],
\end{equation}
the expected all-time maximum of the reflected walk. We note that the stationary waiting time $W_{\infty}$ in the single-server queue with net-input increments $-\xi_{n}$ is distributed as the all-time maximum of the corresponding random walk \cite{Lindley1952}; hence $D(\mu)=\EE(W_{\infty})$ is the mean stationary waiting time of the queue, and for finite-variance increments Kingman's heavy-traffic law mentioned below gives its leading small-$\mu$ asymptotics.

For a Gaussian or strictly stable kernel,
\begin{equation}
\EE(S_{n}^{-}) = c\mku n^{1/\alpha}\hat{h}\Bigl(\frac{\mu n^{\nu}}{c}\Bigr),
\quad
\hat{h}(x) = \EE[(x+Y)^{-}].
\end{equation}
The limiting excess $D(\mu)$ is therefore a harmonic sum with weight $n^{1/\alpha-1}$, and
\begin{equation}
\hat{h}(x) = \int_{x}^{\infty} h(u)\,du,
\quad
\hat{h}'(x) = -h(x).
\end{equation}
If $\hat{h}^{*}(s)=\int_{0}^{\infty} x^{s-1}\hat{h}(x)\,dx$, integration by parts in the common fundamental strip gives
\begin{equation}
\label{eq:shift}
\hat{h}^{*}(s) = \frac{h^{*}(s+1)}{s}.
\end{equation}
Taking the Mellin transform of \eqref{eq:Ddef} then yields
\begin{equation}
\label{eq:Gtilde}
%\begin{split}
\widetilde{G}(s) = c^{1+s}\hat{h}^{*}(s)\zeta(\nu(1+s)) = \frac{c^{1+s}}{s}\,h^{*}(s+1)\zeta(\nu(1+s)).
%\end{split}
\end{equation}
Relative to \eqref{eq:Gmaster}, \eqref{eq:Gtilde} shifts the argument of $h^{*}$ by one unit.

For a Gaussian walk of variance $\sigma^{2}$, the first three poles give
\begin{equation}
\label{eq:Dgauss}
D(\mu) = \frac{\sigma^{2}}{2\mu}+\frac{\zeta(\frac{1}{2})}{\sqrt{2\pi}}\sigma+\frac{\mu}{4}+O(\mu^{2}).
\end{equation}
The first term corresponds to Kingman's heavy-traffic law \cite{Kingman1962}, while the negative (since $\zeta(\frac{1}{2})<0$) constant term is Siegmund's corrected-diffusion constant \cite{Siegmund1979,Siegmund1985,JanssenVanLeeuwaarden2007a}. The two terms thus arise from neighboring poles of $\widetilde{G}$ in \eqref{eq:Gtilde}.

The constant term also connects to the driftless expansion. Writing
\begin{equation}
\EE(M_{n}) = \sigma\sqrt{\frac{2n}{\pi}}-c+o(1),
\end{equation}
the Gaussian value of $c$ (not to be confused with the stable scale parameter used in the rest of the manuscript) is minus the constant term in \eqref{eq:Dgauss}. For a general finite-variance distribution, $c$ contains additional distribution-dependent information. For uniform jumps on $[-1,1]$, for example, $c \simeq 0.29795$ \cite{ComtetMajumdar2005,MajumdarComtet2006,MounaixMajumdar2018}; a Gaussian with the same variance (namely, $\sigma^{2}=1/3$) gives $c \simeq 0.33636$.

For a heavy descending tail, suppose $h(x)\sim C_{-}x^{-\alpha}$. Then
\begin{equation}
\hat{h}(x) \sim \frac{C_{-}}{\alpha-1}\,x^{1-\alpha}.
\end{equation}
Using the scaling form of $\EE(S_{n}^{-})$ above, for each fixed $\mu>0$ we obtain
\begin{equation}
\frac{1}{n}\EE(S_{n}^{-})
\sim \frac{c\mku n^{1/\alpha}}{n}\,
\frac{C_{-}}{\alpha-1}
\Bigl(\frac{\mu n^{\nu}}{c}\Bigr)^{1-\alpha}
= \frac{C_{-}c^{\alpha}}{\alpha-1}\,
\mu^{1-\alpha}n^{1-\alpha},
\quad n \to \infty.
\end{equation}
For $1 < \alpha < 2$ and $C_{-}>0$, the exponent $1-\alpha$ lies in $(-1,0)$, so the nonnegative series in \eqref{eq:Ddef} diverges and $D(\mu)=\infty$. The Mellin representation shows the obstruction directly. The kernel transform requires $\Real{s} < \alpha-1$, whereas the $n$-sum requires $\Real{s}>1/(\alpha-1)$, and the two conditions are incompatible. An infinite excess thus coexists with a positive record rate. At $\beta=+1$, the descending tail is exponentially thin, $C_{-}=0$, the kernel restriction is absent, and $D(\mu)$ is finite.

%% %% %% %% %% %% %% %% %% %% %% %% %% %% %% %% %% %% %% %% %% %% %% %% %% %% %%

\section{Conclusions}

The small-drift record problem separates the information carried by the stable limit from that retained by the increment distribution. Below the crossover $n^{*}$, record accumulation follows the driftless law $N^{\rho}$, while above $n^{*}$ it follows the linear law $\lambda(\mu)N$. Matching the two regimes suggests the exponent $(1-\rho)/\nu$. For Gaussian and strictly stable steps, the double pole of the Mellin transform gives the exponent and prefactor. The correction terms come from the negative-integer poles. The Riemann $\zeta$ factor supplies the harmonic weights in both correction ladders, while the kernel coefficients distinguish the increment distributions. Deterministic evaluations of the Spitzer--Baxter sum agree with the predicted exponent at twenty-seven points across the stable family and with the Gaussian ladder through its cubic term.

Skewness enters through $\rho$. After centering, a heavier ascending tail is accompanied by a left-shifted bulk, which lowers $\rho$ and raises the small-drift exponent. Records are therefore suppressed most strongly at $\beta=+1$ and least strongly at $\beta=-1$. Throughout a stable domain of attraction, $(1-\rho)/\nu$ remains the index of regular variation. The slowly varying multiplier and the lower-order terms retain information about the full increment distribution.

The strictly Cauchy case marks a change of scale. The accumulated shift $\mu n$ and the fluctuations both grow linearly in $n$, and the scaling variable loses its dependence on $n$. The limiting record rate vanishes for every finite location shift, while the shift enters the finite-time exponent $\theta(\mu)$.

For the expected maximum, shifting the kernel transform by one unit produces Kingman's heavy-traffic term and Siegmund's corrected-diffusion constant in the finite-variance case. For $1<\alpha<2$ and a heavy descending tail, the record rate is positive while the limiting excess $D(\mu)$ is infinite. At the light-descending-tail endpoint $\beta=+1$, the excess is finite. Record renewal and the expected maximum thus probe different integrability properties of the one-sided kernel.

For general stable domains of attraction, identifying the slowly varying multiplier and the lower-order corrections requires quantitative information about convergence to the stable law beyond the domain-of-attraction assumption.

%% %% %% %% %% %% %% %% %% %% %% %% %% %% %% %% %% %% %% %% %% %% %% %% %% %% %%

\section*{Acknowledgments}
The author thanks the S\~{a}o Paulo State Research Foundation -- FAPESP, Brazil, for partial support under grant no.~{2020/04475-7}.

%% %% %% %% %% %% %% %% %% %% %% %% %% %% %% %% %% %% %% %% %% %% %% %% %% %% %%

\centerline{$\star$ --- $\star$ --- $\star$}

\end{document}